\documentclass[lettersize,journal]{IEEEtran}
\usepackage{amsmath,amsfonts}
\usepackage{algorithmic}
\usepackage{algorithm}
\usepackage{array}
\usepackage[caption=false,font=normalsize,labelfont=sf,textfont=sf]{subfig}
\usepackage{textcomp}
\usepackage{stfloats}
\usepackage{url}
\usepackage[a4paper,left=1.2cm,right=1.2cm,bottom=1.5cm,top=2.7cm]{geometry}
\usepackage{verbatim}
\usepackage{graphicx}
\usepackage{cite}
\usepackage{xcolor}
\begin{document}
\title{Superdense Coding using Bragg Diffracted Hyperentangled Atoms}

\author{Syed M. Arslan, Saif Al Kuwari, Tasawar Abbas}


\maketitle
\pagestyle{empty}
\begin{abstract}
Superdense coding (SDC) is a popular protocol demonstrating the potential of using quantum mechanics to transfer data, where The sender (Alice) can transfer 2 bits of classical information over a single qubit. We present a scheme for quantum superdense coding through Bragg diffracted hyperentangled atoms generated using cavity quantum electrodynamics (QED). In our scheme, Alice transfers 2 bits of classical information over a single hyperentangled atom. This is achieved by introducing multiple quantum gates using resonant and off-resonant Bragg diffraction in cavity QED setup. This scheme uses multiple degrees of freedom to add an extra layer of security to the encoded information.


\end{abstract}

\section{Introduction} 
\IEEEPARstart{T}{oday}, efficient and secure communication is no longer a luxury. However, while current communication and security systems are often described using the laws of classical physics, it is becoming widely accepted that the laws of quantum mechanics can provide a more accurate description of nature. In fact, the quantum mechanical description of information leads us toward a complete theory of quantum information, paving the way to engineer such systems based on its principles. 

Recent advances in quantum communication have enabled the development of advanced systems such as ultra-secure communication channels \cite{jiang2021quantum} and global quantum networks \cite{anis2021engineering, islam2013generation}, which can harness the full potential of the quantum realm. In fact, the community demonstrated fundamental quantum properties, such as quantum entanglement which acts as a backbone for quantum communication. These demonstrations impact multiple areas of our society, which include security, privacy and networking \cite{10198241,10160024}. 
An emerging field of quantum information is exploiting these fundamental properties to develop new types of information processing and communication which is resistant to the decoherence effect and achievable with modern technologies \cite{10198241}. 

In this paper, we focus on a quantum information processing protocol that effectively transfers two bits of classical information by physically sending one qubit. This protocol is known as quantum superdense coding (SDC). 
 This protocol was first theoretically designed by Bennett and Wiesner \cite{bennett1992communication} in 1992. In 1996, it was experimentally realized by K. Mattle, H. Weinfurter, P. G. Kwiat and A. Zeilinger using two entangled photons \cite{mattle1996dense}. SDC is a quantum communication protocol that provides a stable transfer of classical information over a few qubits, since the communicating parties are already sharing an entangled resource (Bell pair)\cite{harrow2004superdense, liu2002general, abeyesinghe2006optimal}. This communication protocol has
 attracted interest from academia and industry due to its efficiency and its resistance nature to eavesdropping, i.e., to transmit confidential information from one party to another with absolute
security, by exploiting the no-cloning theorem of quantum mechanics \cite{cacciapuoti2019quantum}. However, SDC has not yet reached its full potential. It needs to be implemented in various quantum systems, mainly photonic and atomic systems. Currently, these quantum system have better control of the parameters for its experimental realization. Recently, the systems based on non-linear circuit elements like Josephon Junction are used to realize quantum bits as well as entanglement and further quantum protocols \cite{shahmir2023multi}.

In this paper, we use a Bragg diffracted hyperentangled atom to demonstrate the SDC protocol. In most schemes, entanglement is generated using only one degree of freedom (DOF) among two or more entities. On the other hand, the generation of hyperentanglement is achieved when we exploit multiple degrees of freedom contemporaneously. The initial idea of this special case of entanglement was instigated by P. G. Kwiat \cite{barreiro2005generation}, where the energy and momentum states of a photon were used to generate hyperentanglement. To this date, most researchers have exploited the states of photons to achieve hyperentanglement, but there has recently been a growing interest in manipulating the atomic states to generate hyperentanglement \cite{nawaz2018remote, shi2021hyperentanglement}. 

We utilize the following DOFs: 1) internal energy levels, and 2) external quantized momenta. We show that increasing the number of DOF in entanglement gives an extra layer of security to the encoded data that Alice generates at her end. In our proposed scheme, hyperentangled atoms are engineered using Atomic Bragg Diffraction (ADB), which is a very well-known technique that has been used by the community to solve fundamental problems in quantum information.




\subsection{Contributions}
The contributions of this paper can be summarized as follows:
\begin{itemize}
    \item We proposed a SDC protocol based on hyperentangled atoms and cavity QED
    \item Our proposal demonstrates the creation of quantum gates which are aligned with the atomic hyperentanglment in Bragg regime.
    \item We have provided mathematical calculations that theoretically validates our proposed engineering scheme.
    \item This proposal will provide a framework using multiple DOF, but one can enhance this using decoherence effects.
\end{itemize}

\subsection{Organization}
The rest of this paper is organized as follows: section \ref{ADB} provides a detailed review of Atomic Bragg Diffraction, which includes the historical and mathematical background. Section \ref{hyperSDC}  discusses the implementation of hyperentangled SDC in cavity QED, covering the preparation, sharing, encoding and decoding processes. Finally, sections \ref{experimentalfeas} and \ref{conclusion} discuss the experimental feasibility of the scheme and provide a few concluding remarks. 

\section{Bragg diffraction in Cavity QED} \label{ADB}

In the early 20th century, the discovery of surprising patterns shown by crystal when exposed to an X-ray beam fascinated Lawrence and William Bragg \cite{bragg1913reflection}. They found that when an X-ray beam interacts with crystal at some controlled angle and wavelength, it returns strong peaks of reflected radiation. Further, the incident X-ray radiation will produce a Bragg peak if reflections of the various planes interfere constructively. The constructive interference can only be achieved when the phase angle is a multiple of 2$\pi$. This proposal later evolved, stating that when radiation is incident on a crystal with some angle, radiation will scatter from the lattice and the scattered radiation will interfere constructively \cite{cook1978deflection}. They formalize their findings in a mathematical form known as Bragg's law, $2d\sin\theta=n\lambda$, where $d$ is the separation between the lattice plane, and $\lambda$ is the wavelength of the X-ray radiation.

In the case of atom optics, the roles of the two interacting entities in Bragg diffraction are reversed. In this case, atoms are diffracted from stationary radiation. This special type of stationary radiation is called a cavity field, which can be generated in a controlled environment inside the quantum cavity. A quantum cavity comprises two perfectly reflected mirrors placed opposite each other at some pre-defined distance. The state of the quantized electromagnetic radiation trapped inside the cavity is represented by Fock state $|n\rangle$. Several cavity QED based state engineering schemes can be achieved by choosing the appropriate Fock state inside the cavity \cite{abbas2014state, abbas2015engineering, ali2022hyperentanglement, ali2022teleportation}. When atoms pass through such cavities, they may interact with the Fock state and exchange of atomic momentum may takes place \cite{bernhardt1981coherent}. This momentum exchange between atom and field is based on two factors: 
\begin{enumerate}
    \item Spontaneous Emission Force, when the recoil energy produced is absorbed by the atom and it moves in a random direction.
    \item Dipole Force, when the atom absorbs and emits field quanta.
\end{enumerate}
Due to the negligible atomic lifetime of spontaneous emission force compared to dipole force, we ignore the former \cite{kazantsev1990mechanical}. To enter Bragg's regime, the interaction time between the atom and the field must be large, so that the atom's internal state remains constant. 

\begin{figure}[t]
\centering
\includegraphics[scale=0.6]{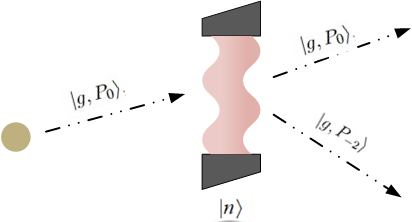}
\caption{Schematic of Bragg Diffraction in cavity QED}
\label{fig:qed}
\end{figure}

Conventional diffraction occurs when light is diffracted by interacting with a solid object (atoms). In our scenario, the roles are reversed; that is, atoms will get diffracted when they interact with a beam of light. This phenomenon is known as Atomic Bragg Diffraction (ADB), which helps investigate fundamental phenomena of quantum physics and gives rise to novel applications\cite{ikram2015wheeler}. The schematics of diffraction is shown in figure \ref{fig:qed}. Here, an atom in its ground state $|g\rangle$ with initial external momenta $|P_0\rangle$ interacts with the quantized EM field which is represented by the Fock state $|n\rangle$. Contrary to classical electromagnetic theory, where energy is continuous in nature, Fock states are discrete energy representations. 

The most basic case of ADB involves a two-level atom in its internal and external states i.e. ground state and its external momenta $|g,P_0\rangle$, which interacts with the cavity field $|n\rangle$. This interaction has two possible outcomes: 1) the atom does not find the field and goes un-interacted, and 2) the atom finds the field and gets deflected with a momentum shift. The probability of the un-interacted scenario is represented by the probability amplitude $C^{P_l}_{n,g}(t)$, while the probability amplitude of the interacted case will be $C^{P_l}_{n-1,e}(t)$ (in the interacted scenario, one photon from the cavity contributes to the deflection of the atom; hence, the number of photons in the cavity will go down to $n-1$ and the atom will become excited $|e\rangle$). This interaction does not satisfy the Bragg condition, as it is not energy conserving; to include it in the Bragg regime, we treat this interaction as the off-resonant case, which prevents the atom's internal state from changing at the end of the interaction, even if it does absorb the photon from the cavity field. The off-resonant condition will force the internal state to complete the cycle and return to its original position. Thus, the only change the atom will experience will be in its momentum state.

We start with a two-level atom, initially in the ground state $|g\rangle$ along with the quantized momenta state $|P_0\rangle$. The cavity is taken as a Fock state $|n\rangle$ as shown in fig [\ref{fig:qed}]. For first-order Bragg diffraction, the momentum gained by the atom from the atom-field interaction is either zero or $2\hbar$. Hence, the quantized atomic momenta for first-order Bragg diffraction are $|P_0\rangle = |\hbar k\rangle $ and $ |P_{-2}\rangle=- |\hbar k\rangle$. The Hamiltonian governing this atom-field interaction under rotating wave approximation is given as follows \cite{islam2008generation, ikram2007engineering}:
\begin{equation}     
    H_i = \frac{P_x^2}{2M}+\frac{\hbar\Delta}{2}\sigma_z+\hbar\mu\cos{kx}[b\sigma_++b^\dag\sigma_-]
\end{equation} \label{eq4}
\noindent where, the first term of the expression $P_x^2/2M$ is the kinetic energy of the system along the x-axis. $P_x$ is the momentum state along x-direction and $M$ is the mass of the particle. $\mu$ represents the coupling constant and $\hbar$ is the planck constant. $b(b^{\dag})$ are field raising and lowering operators. $\sigma_+(\sigma_-)$ are the atomic raising and lowering operators. Finally, $\Delta$ is the detunning in atom-field. The proposed state vector for such condition of an atom is given by:
\begin{align}
     |\psi\rangle= e^{-\iota(\frac{P_0^2}{2M\hbar}-\frac{\Delta}{2})t}\sum \notag\Big[C_{g,n}^{P_l}|g,n,P_l\rangle
     + \\ C_{e,n-1}^{P_l}|e,n-1,P_l\rangle \Big] \label{prop1}
\end{align}
\noindent where, $C_{g,n}^{P_l}(t)[C_{e,n-1}^{P_0}(t)]$ are the probability amplitudes of atom in state $|g\rangle(|e\rangle)$ moving with the generalized atomic momenta $|P_l\rangle$ for $l$ interactions and lastly $|n\rangle[|n-1\rangle]$ is the cavity in Fock state attached to the atom going under interactions. After some rigorous calculations using schr$\Ddot{o}$dinger equation \cite{nawaz2019atomic}, the unknown probability amplitudes are linked forming two coupled differential equations:
\begin{align}
    \frac{\partial}{\partial t} C_{g,n}^{P_l}(t) =  -\iota \Big[\left(\frac{l(l_0+l)\hbar k}{2M}\right)C_{g,n}^{P_l}(t) \label{coup1}
    \\ \notag+ \frac{\mu \sqrt{n}}{2}\left(C_{e,n-1}^{P_{l+1}}(t)+C_{e,n-1}^{P_{l-1}}(t)\right)\Big] 
\end{align}
and
\begin{align}
    \frac{\partial}{\partial t} C_{e,n-1}^{P_l}(t) =  -\iota \Big[\left(\frac{l(l_0+l)\hbar k}{2M}\right)C_{e,n-1}^{P_l}(t) \label{coup2}
    \\ \notag+ \frac{\mu \sqrt{n}}{2}\left(C_{g,n}^{P_{l+1}}(t)+C_{g,n}^{P_{l-1}}(t)\right)\Big] 
\end{align}
As $l$ is a set of $[-\infty,\infty]$, it follows that the coupled differential equations, eq.(\ref{coup1}) and eq.(\ref{coup2}), are infinite coupled equations representing off-resonant and resonant cases, respectively. We will mostly deal with the off-resonant case, hence, accordingly, our detunning $\Delta$ will be very large compared to recoil frequency $\omega$. Solving eq. (\ref{coup1}) and eq.(\ref{coup2}) for first order Bragg diffraction ($l_0=2$), the infinitely large coupled equations cut down to only five coupled equations corresponding to $l \in [-3 ,\cdots 1]$. Applying the off-resonant conditions $\Delta \gg \omega$, then ignoring the $\left(\frac{l(l_0+l)\hbar k}{2M}\right)$ term and applying adiabatic approximation \cite{khan1999quantum} will yield two coupled differential equations:
\begin{align}
 \frac{\partial}{\partial t} C_{g,n}^{P_0}(t) = \iota \left[2\alpha C_{g,n}^{P_0}(t)+ \alpha C_{g,n}^{P_{-2}}(t) \right]
    \\
     \frac{\partial}{\partial t} C_{g,n}^{P_{-2}}(t) = \iota \left[2\alpha C_{g,n}^{P_{-2}}(t)+ \alpha C_{g,n}^{P_{0}}(t) \right] \label{coup4}
\end{align}
where, $\alpha= \mu^2 n/4\Delta$ is the Rabi frequency for n photons. To solve the coupled differential equation, we use Laplace transformation:
\begin{align}
    C_{g,n}^{P_0}(t) = e^{2\iota \alpha t}\left[C_{g,n}^{P_0}(0) \cos({\alpha t}) + \iota C_{g,n}^{P_{-2}}(0) \sin({\alpha t})\right] \label{coupsol1}
    \\
    C_{g,n}^{P_{-2}}(t) = e^{2\iota \alpha t}\left[C_{g,n}^{P_{-2}}(0) \cos({\alpha t}) + \iota C_{g,n}^{P_{0}}(0) \sin({\alpha t})\right] \label{coupsol2}
\end{align}

The expressions in eq. (\ref{coupsol1}) and eq. (\ref{coupsol2}) are the fundamental tools for engineering quantum optics tools. For instance, if we want to engineer an atomic beamsplitter, we set the interaction time $\alpha t = \pi/4$, and similarly, the interaction time $\alpha t = \pi/2$ for the atomic mirror in the QED scenario of the cavity \cite{ikram2015wheeler}. 

\section{Hyperentangled SDC in Cavity QED}\label{hyperSDC}
SDC involves two remote parties possessing entangled qubits, traditionally called Alice and Bob, who want to transfer a secret string of classical bits. Alice generates a new qubit and hides the \emph{secret string} in her qubit using basic quantum operators; this process is known as encoding. The encoded qubit of Alice is sent to Bob via a quantum channel \cite{anis2021engineering,islam2013generation}. 
Bob retrieves the classical bits sent by Alice by applying various operations on both the qubit he receives from Alice and the entangled qubit he already has. 

In this section, we will show how SDC can be implemented using hyperentangled atoms and cavity QED. We will discuss encoding (by Alice), sharing (by Alice) and decoding (by Bob).


\begin{figure}
\centering
\includegraphics[scale=0.6]{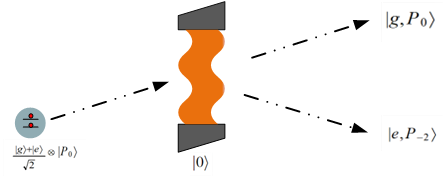}
\caption{This figure represents the schematics of generation of hyperentangled atom}
\label{fig:hypsup}
\end{figure}

    
\subsection{Preparation} \label{hypsup}
To execute the SDC protocol, a preparation phase needs to be completed. This phase involves the generation of hyperentangled atoms and creating the Bell pairs. 


\subsubsection{Generating Hyperentangled atoms} \label{hypatom}
We first present the scheme for generating a hypersuperposition state between the atomic internal and external degrees of freedom. A two-level atom that is prepared in a superposition of its internal states ${|g\rangle+|e\rangle}/2$, where the external momenta state $|P_0\rangle$ interacts with the cavity in the vacuum state $|0\rangle$. Hence, the initial state vector is: 
\begin{equation} 
     |\psi\rangle=\frac{1}{\sqrt{2}} [|g\rangle+|e\rangle]\otimes|P_0\rangle\otimes|0\rangle\label{eq2}
        \end{equation}
The interaction between the particle and the field follows the same procedure as in \ref{ADB}, but with a slight change. In our scheme, the initial state of the atom is in a superposition of its internal state, and the cavity with which it has to interact is in a vacuum state. In addition, the interaction discussed in section \ref{ADB}  was under the off-resonant condition, whereas this interaction will be under the resonant condition. As the atom here will emanate an extra term of the $|e\rangle$ state, the proposed state vector will become:
\begin{align}
    |\psi\rangle= e^{-\iota(\frac{P_0^2}{2M\hbar}-\frac{\Delta}{2})t}\sum \Big[C_{g,0}^{P_0}|g,0,P_l\rangle \notag
    \\+C_{g,1}^{P_l}|g,1,P_l\rangle+C_{e,0}^{P_l}|e,0,P_l\rangle \Big]  \label{eq3}
\end{align}
The interaction Hamiltonian governing this scheme under the rotating wave and dipole approximation is the same as in section \ref{ADB}. Using the Schr$\Ddot{o}$dinger equation and the Hamiltonian stated in section \ref{ADB}, yields the following set of differential equations:
\begin{align}
     \iota \Dot{C}_{g,0}^{P_l}(t) &= 0 \label{eq5} \\
     \iota \Dot{C}_{e,0}^{P_l}(t) &= \left[\frac{l(l_0+l)\hbar k^2}{2M}\right] C_{e,0}^{P_l}(t)+ \frac{\mu}{2}\left[C_{g,1}^{P_{l+1}}(t)+ C_{g,1}^{P_{l-1}}(t)\right]\label{eq6}\\
     \iota \Dot{C}_{g,1}^{P_l}(t) &= \left[\frac{l(l_0+l)\hbar k^2}{2M}\right] C_{g,1}^{P_l}(t)+ \frac{\mu}{2}\left[C_{e,0}^{P_{l+1}}(t)+ C_{e,0}^{P_{l-1}}(t)\right]\label{eq7}
\end{align}
These differential equations represent the rate of change of the probability amplitudes. The first expression in eq (\ref{eq5}) tells us that it is working independently of the other equations and will generate a common solution based on the initial conditions. The other two equations, (\ref{eq6}) and (\ref{eq7}), are an infinite set of coupled differential equations. According to first-order Bragg conditions, $l_0=2$, assuming that significant contributions come from the values of $l$ ranging from [-3 to 1]. Substituting these values in eq (\ref{eq6}) and eq (\ref{eq7}), we obtain:
\begin{equation}
      \iota \Dot{C}_{e,0}^{P_{-2}}(t) = \frac{-\iota\beta}{2} \left[\frac{1}{2}C_{e,0}^{P_0}(t)+\frac{1}{3}C_{e,0}^{P_{-2}}(t)\right]\label{eq8}
\end{equation}
\begin{equation}
       \iota \Dot{C}_{e,0}^{P_{0}}(t) = \frac{-\iota\beta}{2} \left[\frac{1}{2}C_{e,0}^{P_{-2}}(t)+\frac{1}{3}C_{e,0}^{P_0}(t)\right]\label{eq9}
\end{equation}
where the constant $\beta=\mu^{2}/w_r$ and $w_r=\hbar k^2/2M$. We observe that the above differential equations are coupled with each other, which can only happen using the Laplace transformation. As a result, we get the solution of the rate of change of the probability amplitude as:
\begin{equation}
       C_{e,0}^{P_{0}}(t) = e^{-\iota\frac{\beta}{6}t} \left[\cos\left({\frac{\beta}{4}t}\right)C_{e,0}^{P_{0}}(0)-\iota\sin\left({\frac{\beta}{4}t}\right) C_{e,0}^{P_{-2}}(0)\right] \label{eq10}
\end{equation}
\begin{equation}
      C_{e,0}^{P_{-2}}(t) = e^{-\iota\frac{\beta}{6}t} \left[\cos\left({\frac{\beta}{4}t}\right)C_{e,0}^{P_{-2}}(0)-\iota\sin\left({\frac{\beta}{4}t}\right) C_{e,0}^{P_{0}}(0)\right] \label{eq11}
\end{equation}
Substituting the initial conditions from the initial atom ($C_{g,0}^{P_{0}}(0)=1/\sqrt{2}$,  $C_{e,0}^{P_{0}}(0)=1/\sqrt{2}$ and $C_{e,0}^{P_{-2}}(0)=0$) to eq. (\ref{eq5}), eq. (\ref{eq10}) and eq. (\ref{eq11}) along with the fixed interaction time $t=2\pi/\beta$, gives us the final state of the atom that emerges from the cavity as in \cite{nawaz2017engineering}:
\begin{equation}
    |\psi\rangle = \frac{1}{\sqrt{2}}\left(|g,P_{0}\rangle - \iota |e, P_{-2}\rangle\right) \label{eq12} 
\end{equation}
The state in eq. (\ref{eq12}) represent the atom in a superposition state of not only its internal state, but also its external momenta simultaneously. We also note that the cavity state was the same $|0\rangle$ throughout the calculations, which is not shown in all previous expressions.
\begin{figure}
    \centering
    \includegraphics[scale=0.5]{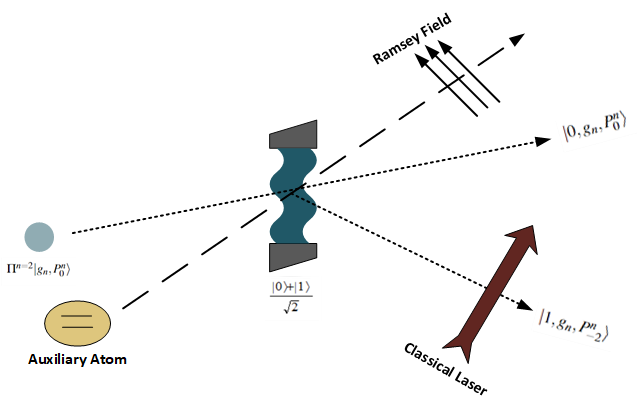}
    \caption{Charlie's end: Schematic for the generation of hyperentangled Bell state \cite{nawaz2017engineering}}
    \label{fig:Bell}
\end{figure}

\subsubsection{Generation of Bell pairs} \label{bell}
According to SDC, the two parties, 
Alice and Bob, need to establish some quantum link with each other. To create this link, we have to create hyperentangled Bell pairs (hyperentangled atoms) and share them with the parties before initiating the protocol. The building block of the Bell pair has been laid in section \ref{hypatom} by generating an atom that is prone to be entangled in multiple degrees of freedom.

The generation of hyperentangled Bell state is slightly different from the scheme presented in section \ref{hypatom}. We will consider off-resonant conditions and, this time, we will prepare the cavity in the superposition state $(|0\rangle+|1\rangle)/\sqrt{2}$. There will be two neutral atoms initially prepared in the ground state $|g\rangle$ and in the atomic momenta $|P_0\rangle$. The atoms will pass through the cavity off-resonantly, one after another, resulting in the initial state after the first atom interacts:
\begin{align}
     |\psi_1\rangle = |g_1,P_{0}^{(1)}\rangle \otimes (|0\rangle+|1\rangle)/\sqrt{2} \label{eq13}
\end{align}
After taking all the considerations presented above, the proposed state vector of the first interaction will be given as:
\begin{align}
    |\psi\rangle= e^{-\iota(\frac{P_0^2}{2Mh}-\frac{\Delta}{2})}\sum \Big[C_{g_1,0}^{P_0^{(1)}}(t)|g_1,0,P_{l}^{(1)}\rangle \notag
    \\+C_{g_1,1}^{P_l^{(1)}}(t)|g_1,1,P_{l}^{(1)}\rangle+C_{e_1,0}^{P_{l}^{(1)}}(t)|e_1,0,P_{l}^{(1)}\rangle \Big] \label{eq14}
\end{align}

The Hamiltonian of the system is the same as in eq (\ref{eq4}). Solving the system using Schr$\Ddot{o}$dinger equation and substituting the values of the initial condition and interaction time $t_1=2\Delta_1\pi/\beta$, the final state of the first interaction becomes:
\begin{equation}
    |\psi_1\rangle = \frac{1}{\sqrt{2}} \left(|g_1,0,P_{0}^{(1)}\rangle - \iota |g_1,1,P_{-2}^{(1)}\rangle\right) \label{eq15}
\end{equation}
We note that there is no hyperentanglement generated yet, the only entanglement generated so far is between the 
external momenta state and the cavity field. Similarly, the second atom $|g_2,P_{0}^{(2)}\rangle$ is released, which will interact with the same cavity. In this case, the cavity is the same but with a different initial state. Expression (\ref{eq15}) will serve as the initial condition for the following interaction. Using the interaction time $t_2 = 2\Delta_2\pi/\beta$ the new state becomes:
\begin{equation}
    |\psi_{1,2}\rangle = \frac{1}{\sqrt{2}} \left(|g_1,g_2,0,P_{0}^{(1)},P_{0}^{(2)}\rangle - |g_1,g_2,1,P_{-2}^{(1)},P_{-2}^{(2)}\rangle\right) \label{eq16}
\end{equation}
Eq. (\ref{eq16}) is a Bell state between an atom and a cavity, but to make the two atoms hyperentangled with each other we have to go through two steps. First, we introduce a two-level auxiliary atom in the ground state $|g_s\rangle$ interacting with the cavity resonantly. This auxiliary atom acts as a catalyst and its main objective is to swap the entanglement of atom-cavity to atom-atom. At the end, the auxiliary atom is traced out of the system by detection. The Hamiltonian used for this interaction is $H_i = \hbar\mu^{(s)}\left(\sigma_+^{(s)}a+\sigma_-^{(s)}a^{\dag}\right)$ and the initial state is $1/\sqrt{2}\left(|g_1,g_2,0,P_{0}^{(1)},P_{0}^{(2)}\rangle - \iota |g_1,g_2,1,P_{-2}^{(1)},P_{-2}^{(2)}\rangle\right)\otimes|g_s\rangle$, where $\sigma_+$ and $\sigma_-$ are the raising and lowering operators. This step swaps the entanglement from the cavity to the auxiliary atom. The new state is: 
\begin{align}
    |\psi_{1,2,s}\rangle = \frac{1}{\sqrt{2}} \Big(|g_1,g_2,g_s,P_{0}^{(1)},P_{0}^{(2)}\rangle \notag \\ - \iota |g_1,g_2,e_s,P_{-2}^{(1)},P_{-2}^{(2)}\rangle \Big) \otimes |0\rangle \label{eq17}
\end{align} 

The un-entangled cavity state $|0\rangle$ is duly traced from the above expression. To dispose of the auxiliary atom, we pass the atom through a Ramsey field. The operation of the Ramsey field is a carbon copy of a Hadamard gate i.e. $|g_s\rangle \rightarrow (|g_s\rangle+|e_s\rangle)/\sqrt{2}$ and $|e_s\rangle \rightarrow (|g_s\rangle-|e_s\rangle)/\sqrt{2}$. After the operation of the Ramsey field, the auxiliary atom is traced out of the system, creating hyperentanglement among the two initial atoms. The eq (\ref{eq17}) transforms into:

\begin{align}     
    |\psi_{1,2}\rangle = \frac{1}{\sqrt{2}}\Big[ \Big(|g_1,g_2,P_{0}^{(1)},P_{0}^{(2)}\rangle - \iota |g_1,g_2,P_{-2}^{(1)},P_{-2}^{(2)}\rangle \Big) \notag \\
    \otimes |g_s\rangle +
  \Big(|g_1,g_2,P_{0}^{(1)},P_{0}^{(2)}\rangle + \iota |g_1,g_2,P_{-2}^{(1)},P_{-2}^{(2)}\rangle \Big)  \otimes|e_s\rangle \Big]\label{eq18}
\end{align}

This expression demonstrates how the entanglement is swapped from the auxiliary atom to the atoms of Alice and Bob. The Bell state produced here depends on the detection of the auxiliary atom. For a simpler case we detect our system in the ground state of auxiliary atom $|g_s\rangle$. Thus, the expression in (\ref{eq18}) becomes:
\begin{align}
    |\psi_{bell}\rangle = |g_1,g_2,P_{0}^{(1)},P_{0}^{(2)}\rangle - \iota |g_1,g_2,P_{-2}^{(1)},P_{-2}^{(2)}\rangle \label{hypstate}
\end{align}

\begin{figure*}[t]
\centering
\includegraphics[scale=0.25]{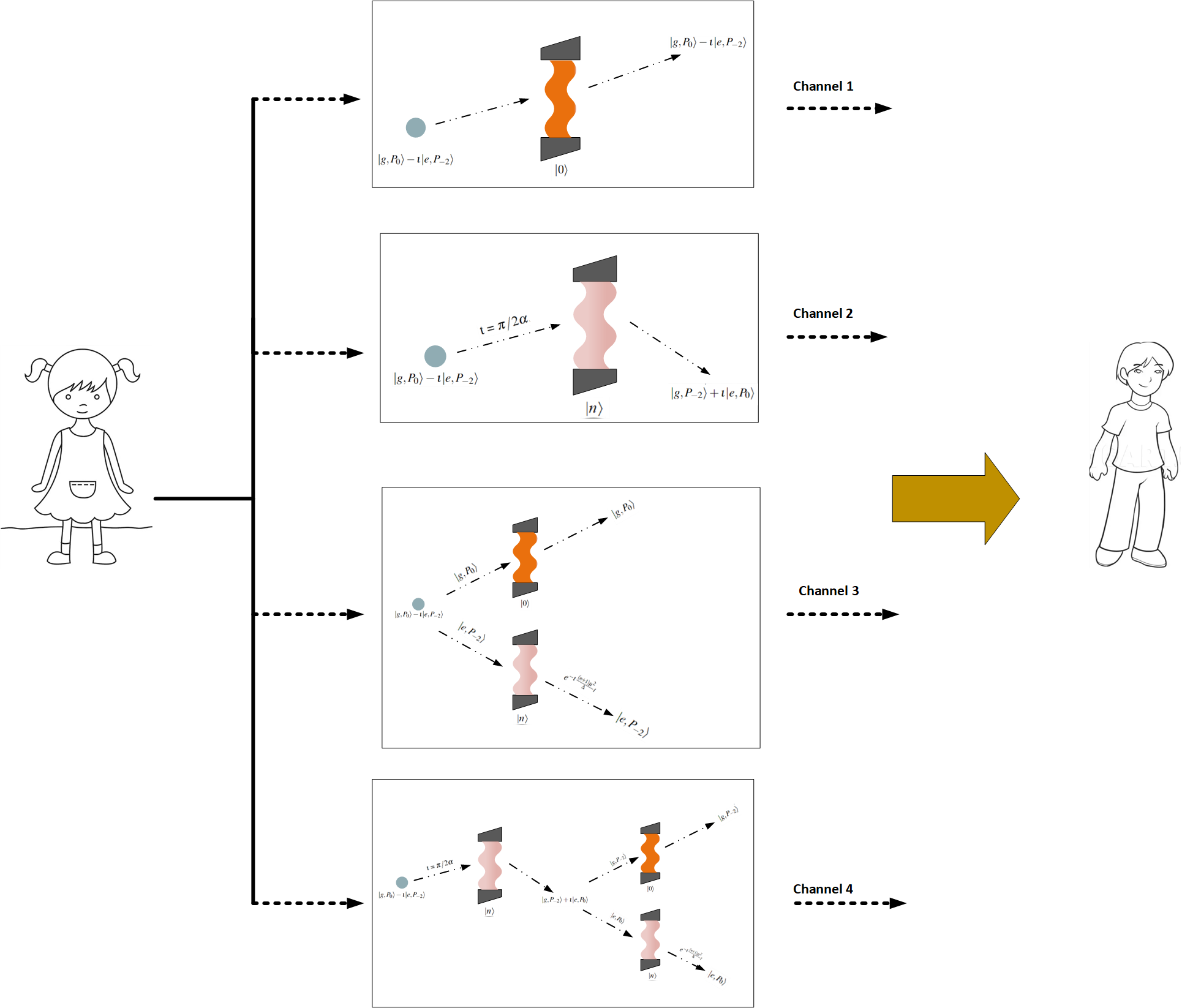}
\caption{The encoding process at Alice's end. She has four channels from which she can send her desired message. If she chooses channel 1, she encodes `00', channel 2 encodes `01', channel 3 encodes `10' and channel 4 will encode `11'}

\label{fig:AliceEnd}
\end{figure*}

Now, we have successfully generated a Bell state but it is not in the hyperentangled bell state. As shown in the expression (\ref{hypstate}), only the momenta states of Alice's and Bob's atoms are correlated, which makes this entanglement one degree of freedom. To unlock multiple degrees of freedom, i.e. hyperentanglement, we pass one arm of the state $\left(|g_1,g_2,P_{-2}^{(1)},P_{-2}^{(2)}\rangle\right)$ through a classical laser. This interaction is governed by a semi-classical Hamiltonian $H= \hbar\Omega_{r}/2[e^{-\iota\phi}\sigma_+ + e^{\iota\phi}\sigma_-]$ \cite{scully1999quantum}, which effectively excites the internal states of the atoms from $|g\rangle$ to $|e\rangle$ and hence generates our required hyperentangled Bell state:
\begin{equation}
      |\psi_{bell}\rangle = |g_1,g_2,P_{0}^{(1)},P_{0}^{(2)}\rangle - \iota |e_1,e_2,P_{-2}^{(1)},P_{-2}^{(2)}\rangle \label{eq20}
\end{equation}
As we can now see, the internal states and the external states of both atoms are entangled simultaneously.

\subsection{Sharing}
Once the hyperentangled Bell state is prepared in eq (\ref{eq20}), we write the equation to emphasize Alice and Bob's atoms as follows:
\begin{equation}
     |\psi_{bell}\rangle = |g_A,g_B,P_{0}^{(A)},P_{0}^{(B)}\rangle - \iota |e_A,e_B,P_{-2}^{(A)},P_{-2}^{(B)}\rangle \label{eq21}
\end{equation}
where the subscript for the first atom ``A'', denotes Alice's atom, and the second ``B" denotes Bob's atom.
As this state is prepared by Charlie, he can send these atoms to Alice and Bob. The atom denoted by subscript ``A'' goes to Alice and ``B'' goes to Bob.

As now the quantum link has been established between the two parties, Alice can proceed with the transmission of her classical information to Bob using her Bell state qubit, by applying different gates.

\subsection{Encoding}

At this point, Alice has received her part of the Bell state from Charlie and wishes to utilize the quantum channel created with Bob to send some classical bits. To achieve this, she must manipulate her existing hyperentangled Bell state through various quantum gates. As her system is in an atomic state, she needs to engineer these quantum gates using a cavity QED with pre-determined initial conditions. The four possible Bell states corresponding to their respective classical bits are shown in the table below (for brevity, we ignore the subscript ``A'' in this section as all manipulations take place on Alice's end):

\begin{table}[H]
    \centering
  \caption{The table shows all possible hyperentangled Bell states corresponding to one hyperentangled atom}
    \begin{tabular}{l|c|c}
       $\beta_{x,y}$  & Bell state/$\sqrt{2}$ & Intended message \\ \hline
       $\beta_{00}$  & $|g,P_{0}\rangle-\iota|e,P_{-2}\rangle$ & 00 \\
       $\beta_{01}$ & $|g,P_{-2}\rangle-\iota|e,P_{0}\rangle$ & 01 \\
       $\beta_{10}$ & $|g,P_{0}\rangle-\iota e^{\iota\alpha}|e,P_{-2}\rangle$ & 10 \\
       $\beta_{11}$ & $|g,P_{-2}\rangle-\iota e^{\iota\alpha}|e,P_{0}\rangle$ & 11
    \end{tabular}
    \label{Belltable}
\end{table}
Table \ref{Belltable} depicts the correspondence between the classical messages and the quantum mechanical state of the sender's system. Whenever Alice intends to send the classical message 00, she does not need to apply any operation to her existing Bell state, whereas she needs to apply an additional gate operations if she intends to send other messages: 01, 10, 11.

In this section, we discuss the necessary conditions for a cavity QED, which can generate outputs that match the various Bell states in Table \ref{Belltable}. In particular, We discuss four gates: Identity, Not, Phase and Phase-Not gates.

\subsubsection{Identity Gate}
This gate is straightforward as we only need to pass the initial state in its original form as the output. To engineer the identity gate in cavity QED, we take our cavity in the vacuum state $|0\rangle$. As Alice atom is in the initial state $|g,P_0\rangle - \iota |e,P_{-2}\rangle$, the external momenta states will pass undisturbed since there is no photon in the cavity to interact with. However, the output will be the same as the input, as shown in fig (\ref{fig:gates}), where we can see that classical bit 00 has been encoded successfully in the form of $\beta_{00}$ Bell state.
\begin{equation*}    
|\beta_{00}\rangle_{in} = |g,P_0\rangle - \iota |e,P_{-2}\rangle \rightarrow |g,P_0\rangle - \iota |e,P_{-2}\rangle = |\beta_{00}\rangle_{out}
\end{equation*}

\subsubsection{Not Gate}
The Not gate is common in atomic Bragg diffraction, where it functions identically to an atomic mirror. The setup of this gate requires a cavity in the Fock state $|n\rangle$ along with a pre-determined interaction time of $t = \pi/2\alpha$, where $\alpha = \mu^2/4\Delta$. When the atomic momenta state interacts with the cavity using these conditions, it effectively flips the input momenta state, giving a phase with it. The input state $|P_0\rangle$ goes to $\iota |P_{-2}\rangle$ and $|P_{-2}\rangle$ goes to $\iota |P_{0}\rangle$. The internal states remain unchanged. The process is explained in the expression below:
\begin{equation*}     
|\beta_{00}\rangle_{in} = |g,P_0\rangle - \iota |e,P_{-2}\rangle \rightarrow |g,P_
{-2}\rangle - \iota |e,P_{0}\rangle = |\beta_{01}\rangle_{out}
\end{equation*}

In this case, Alice has encoded the classical bit of 01 in the form $\beta_{01}$ Bell state.

\subsubsection{Phase Gate}
\begin{figure*}[h]
\centering
\includegraphics[scale=0.4]{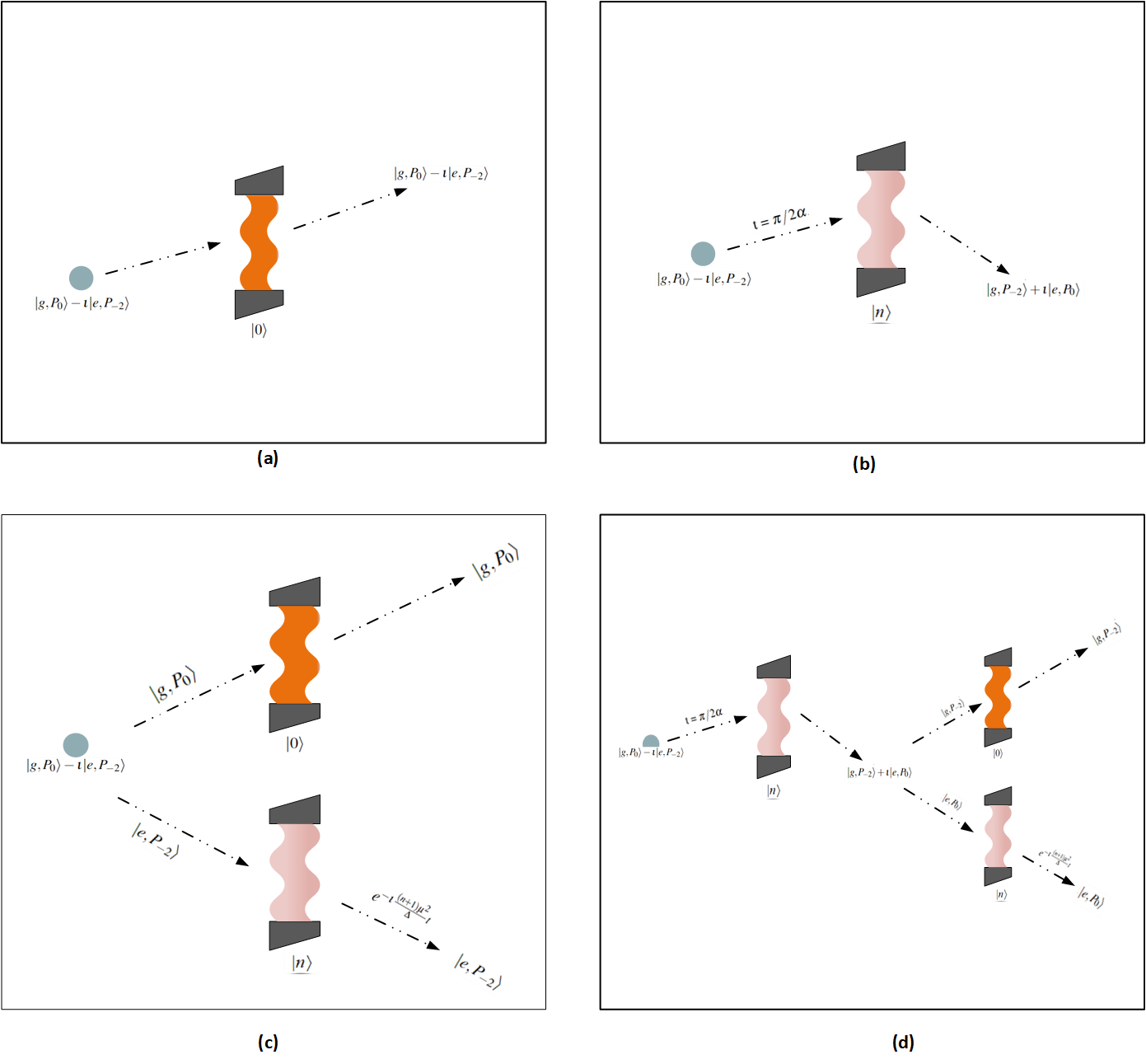}
\caption{Illustration of gates: (a) represents the identity gate, (b) represents the NOT gate, (c) represents the Phase gate, (d) represents the Phase-NOT gate}
\label{fig:gates}
\end{figure*}

In the computational basis, the phase gate has no effect on the input state $|0\rangle$ while it incorporates a phase with input state $|1\rangle$. To mirror this scenario in cavity QED, we take one cavity in Fock state $|n\rangle$ interacting with one arm of the atomic state $|e,P_{-2}\rangle$ off-resonantly. The Hamiltonian describing this interaction is given by $H_{eff}= \hbar\mu^2/\Delta(aa^{\dag}|e\rangle\langle e|- a^{\dag}a|g\rangle\langle g|)$, where $\mu$ is the coupling constant, and $\Delta$ is the detunning and $aa^{\dag}$ are the lowering and raising field operators. A general time evolved state arises after the Hamiltonian is applied along with the necessary conditions $|\psi (t)\rangle = e^{-\iota(n+1)\mu^2 t/\Delta} |\psi (0)\rangle$. This time evolved state shows that every interaction will be incorporated with a phase depending on the initial condition of the state. In this case, one arm of the atom $(|g,P_{0}\rangle)$ is not interacting with the cavity as shown in the figure \ref{fig:gates}, so it will not change in the end. On the other hand, the interacting arm $|e,P_{-2}\rangle$ will get the phase from the cavity and the state will not change in the end $e^{-\iota(n+1)\mu^2 t/\Delta} |e,P_{-2}\rangle$. The process from input to output is shown below:
\begin{align}
\begin{split}
    |\beta_{00}\rangle_{in} = |g,P_0\rangle - \iota |e,P_{-2}\rangle \notag
     \rightarrow 
     \\|g,P_{0}\rangle - \iota e^{-\iota\frac{(n+1)\mu^2}{\Delta}t}|e,P_{-2}\rangle  = |\beta_{10}\rangle_{out}
     \end{split}
\end{align}

In this case, Alice has successfully encoded 10 classical bits in the Bell state $\beta_{10}$.

\subsubsection{Phase-NOT Gate}
Phase-Not gate is a special gate composed of two gates (Not and Phase) at the same time. It flips the input state and embodies the phase into the state $|1\rangle$. We assemble our cavity QED setup in such a way that we flip the momenta state of the input qubit $|g,P_0\rangle-\iota|e,P_{-2}\rangle$ and add a phase to one arm of the state $|e,P_{-2}\rangle$. This gate is implemented in two steps. First, we set the cavity in the Fock state $|n\rangle$ along with the interaction time $t = \pi/2\alpha$ under Bragg conditions. The transformed state after this interaction is given as follows:
\begin{equation*}     
|\beta_{00}\rangle_{in} = |g,P_0\rangle - \iota |e,P_{-2}\rangle \rightarrow |g,P_
{-2}\rangle - \iota |e,P_{0}\rangle
\end{equation*}
The second step involves one arm of the above mentioned state $|e,P_{0}\rangle$ to interact with another cavity that is in Fock state off-resonantly. The Hamiltonian governing this step is $H_{eff}= \hbar\mu^2/\Delta(aa^{\dag}|e\rangle\langle e|- a^{\dag}a|g\rangle\langle g|)$. The final state after both steps is given as:
\begin{equation*}
    |g,P_
{-2}\rangle - \iota |e,P_{0}\rangle \rightarrow |g,P_{-2}\rangle - \iota e^{-\iota\frac{(n+1)\mu^2}{\Delta}t}|e,P_{0}\rangle = |\beta_{11}\rangle_{out}
\end{equation*}
In this case, Alice has encoded her qubit to $\beta_{11}$, which represents the classical information of 11.



\subsection{Decoding}
Once Alice successfully encodes her (classical) message in her qubit, she sends it to Bob via some quantum channel. 
On the other side, Bob operates on the state of two entangled pair (the qubit he just received from Alice and his own qubit), which leads to one of the four Bell states given in Table \ref{Belltable} along with Bob's state. The four combined Bell states are given as:
\begin{table}[H]
    \centering
\caption{All possible hyperentangled Bell states corresponding to Alice and Bob hyperentangled atoms}
   \scalebox{0.9}{ \begin{tabular}{l|c|c} 
       $\beta_{x,y}$  & Bell state/$\sqrt{2}$ &  message \\ \hline
       $\beta_{00}$  & $|g_A,g_B,P_{0}^{A},P_{0}^{B}\rangle-\iota|e_A,e_B,P_{-2}^{A},P_{-2}^{B}\rangle$ & 00 \\
       $\beta_{01}$ & $|g_A,g_B,P_{-2}^{A},P_{0}^{B}\rangle-\iota|e_A,e_B,P_{0}^{A},P_{-2}^{B}\rangle$ & 01 \\
       $\beta_{10}$ & $|g_A,g_B,P_{0}^{A},P_{0}^{B}\rangle-\iota e^{\iota\alpha}|e_A,e_B,P_{-2}^{A},P_{-2}^{B}\rangle$ & 10 \\
       $\beta_{11}$ & $|g_A,g_B,P_{-2}^{A},P_{0}^{B}\rangle-\iota e^{\iota\alpha}|e_A,e_B,P_{0}^{A},P_{-2}^{B}\rangle$ & 11
    \end{tabular}}
    \label{Belltable2}
\end{table}


Now, let us assume that Alice wants to send the classical bits 01. For that, Alice uses the channel that encodes her Bell state as $|\beta_{01}\rangle$. Now, Bob has a combined state mentioned in Table (\ref{Belltable2}). To retrieve data from this encoded state, Bob sends this state through two black boxes, which are aligned one after another. The first black box is responsible for removing the extra layer of security provided by the hyperentanglement, whereas the second black box is responsible for converting the quantum state into classical information. The following section will discuss these black boxes in detail.

\subsubsection{BlackBox 1: Removing extra layer of security}
As it has been established previously that the pre-entangled pair in this scheme is a special kind of entanglement known as hyperentanglement. Therefore, the decoding process would be proportional to the number of degrees at which the atoms are entangled simultaneously. In this very case the atoms are entangled in two degrees of freedom simultaneously i.e. internal energy states and external atomic momenta (see sec: (\ref{hypsup})). Here we will get rid of the internal energy states which we consider as the extra security layer on the encoded bits. Now continuing with the assumption that Alice wants to send `01', the initial becomes:
\begin{equation}
     |\beta_{01}\rangle =  \frac{1}{\sqrt{2}} \left[|g_{A},g_{B},P_{-2}^{A},P_0^{B}\rangle - \iota |e_{A},e_{B},P_{0}^{A},P_{-2}^{B}\rangle \right]
\end{equation}
Inside Blackbox 1, at first, Alice's state experiences a Ramsey field (see sec:(\ref{bell})). The initial state will be transformed into:
\begin{equation}
\begin{aligned}
    |\beta_{01}\rangle= \frac{1}{2\sqrt{2}}\Big{[}|g_{A},g_{B},P_{-2}^{A},P_0^{B}\rangle  +|e_{A},g_{B},P_{-2}^{A},P_0^{B}\rangle-\\ \iota |g_{A},e_{B},P_{0}^{A},P_{-2}^{B}\rangle + \iota|e_{A},e_{B},P_{0}^{A},P_{-2}^{B}\rangle \Big{]}
    \end{aligned} \label{AliceRAM}
\end{equation}
The expression \ref{AliceRAM} is again exposed to Ramsey field, but this time Bob's states get affected. The new combined state is given as:
\begin{align}
       |\beta_{01}\rangle=  &\frac{1}{4\sqrt{2}}\Big{[}|g_{A},g_{B},P_{-2}^{A},P_0^{B}\rangle
       +
       |g_{A},e_{B},P_{-2}^{A},P_0^{B}\rangle \label{BOBRAM}
         \\ \notag &+|e_{A},g_{B},P_{-2}^{A},P_0^{B}\rangle 
       +|e_{A},e_{B},P_{-2}^{A},P_0^{B}\rangle
       \\ \notag & - \iota |g_{A},e_{B},P_{0}^{A},P_{-2}^{B}\rangle +\iota|g_{A},g_{B},P_{0}^{A},P_{-2}^{B}\rangle \\ \notag &- \iota|e_{A},g_{B},P_{0}^{A},P_{-2}^{B}\rangle + 
       \iota|e_{A},e_{B},P_{0}^{A},P_{-2}^{B}\rangle \Big{]}    
 \end{align}

The expression \ref{BOBRAM} can be rearranged by taking out the internal states of the hyperentangled atom. The expression becomes:
\begin{equation}
    \begin{split}
     |\beta_{01}\rangle=\frac{1}{4\sqrt{2}}\Big[|P_{-2}^{A},P_0^{B}\rangle + \iota |P_{0}^{A},P_{-2}^{B}\rangle \otimes |g_A, g_B\rangle \\ +
     |P_{-2}^{A},P_0^{B}\rangle - \iota |P_{0}^{A},P_{-2}^{B}\rangle \otimes |g_A, e_B\rangle \\ + 
    |P_{-2}^{A},P_0^{B}\rangle + \iota |P_{0}^{A},P_{-2}^{B}\rangle \otimes |e_A, g_B\rangle \\+
    |P_{-2}^{A},P_0^{B}\rangle - \iota |P_{0}^{A},P_{-2}^{B}\rangle \otimes |e_A, e_B\rangle \Big]
    \end{split}
\end{equation}

As shown in the above expression, internal energy states are tensor-ed out of the Bell state. Therefore, we can now say that the hyperentanglement is completed and the state becomes normally entangled. Bob detects the state in $|g_A,e_B\rangle$, leaving him with the Bell state $|P_{-2}^{A},P_0^{B}\rangle - \iota |P_{0}^{A},P_{-2}^{B}\rangle$. Now Bob will proceed with the extraction process from the Bell state obtained. The extraction process takes place in Blackbox 2.

\begin{figure*}[t]
\centering
\includegraphics[scale=0.4]{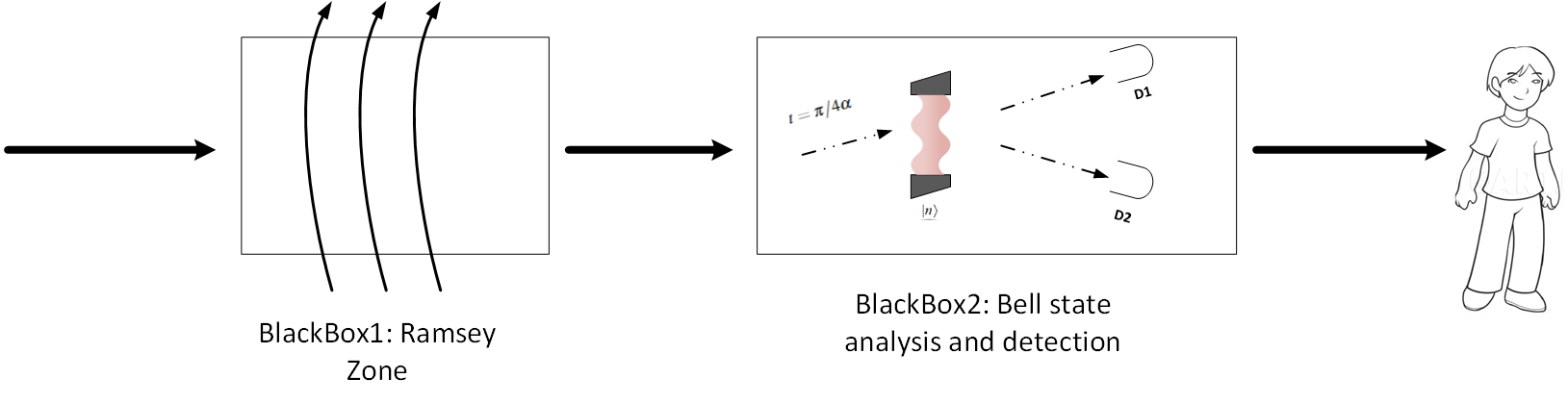}
\caption{The decoding process at Bob's end. In Blackbox 1, the state received from Alice goes through a Ramsey field and is passed on to Blackbox 2. In Blackbox 2, the state interacts with an atomic beamsplitter and states are detected at the end, which are observed by Bob.}
\label{fig:BobEnd}
\end{figure*}

\subsubsection{BlackBox 2: Information extraction process}
Bob uses the Bell state measurement technique to extract classical information from the entangled pair. He first sends the Bell state $|P_{-2}^{A},P_{0}^{B}\rangle-\iota|P_{0}^{A},P_{-2}^{B}\rangle$ to a cavity that is in Fock state along with the interaction time $\tau= \pi/4\beta$ (see fig \ref{fig:BobEnd}), where the cavity is prepared as an atomic beamsplitter, as discussed in section \ref{ADB}. The interaction of both atoms happens simultaneously. The combined state of Alice and Bob transforms into:
\begin{equation}
\begin{split}
    |P_{-2}^{A}\rangle \rightarrow |P_{-2}^{A}\rangle - \iota |P_0^{A}\rangle
    \\
    |P_{0}^{A}\rangle \rightarrow |P_{0}^{A}\rangle - \iota |P_{-2}^{A}\rangle
    \\
    |P_{0}^{B}\rangle \rightarrow |P_{0}^{B}\rangle - \iota |P_{-2}^{B}\rangle
     \\
    |P_{-2}^{B}\rangle \rightarrow |P_{-2}^{B}\rangle - \iota |P_{0}^{B}\rangle
    \end{split}
\end{equation}
After some mathematical operations, the Bell state collapses to $2\iota|P_{0}^{A},P_{-2}^{B}\rangle$. Finally, the detectors detect the quantum state and we are left with the combination of the atoms of Alice and Bob $P_{0}^{A}P_{-2}^{B}$. This means that Alice wanted to send the classical message 01 to Bob, as the $|P_0\rangle$ and $|P_{-2}\rangle$ momenta state of the atom represent 0 and 1 in computational states, respectively. This process is valid for the other three Bell bases, too; hence, if Alice wants to send a long message, she can repeat the exercise multiple times.  Measuring the Bell states at Bob's site efficiently is a crucial aspect of quantum information processing, which requires far more superior detectors and highly controlled environment. However, the Hyperentangled Bell State Analysis (HBSA) has been realized using various protocols in recent times 
\cite{walborn2003hyperentanglement}, \cite{williams2017superdense}.

\section{Experimental feasibility} \label{experimentalfeas}
Cavity QED is a common technique that has been widely used to perform many tasks related to quantum information, such as generating quantum entanglement or creating quantum channels \cite{haroche2006exploring}. In recent times, major improvements can be seen in the lifetimes of the cavities. Due to this advancement, there is a major reduction in decoherence risks, which arise due to spontaneous emissions. The system chosen in this process is in a closed environment where noise and other lossy factors have been ignored. Nevertheless, cavities life-times and atom-field interaction times are setup within the real boundaries, which makes our closed system approximation justified and robust to the decoherence threats \cite{park2022slowing}, \cite{walborn2003hyperentanglement}, \cite{wang2019complete} \cite{gianni2021new}. As far as Bragg regime atomic diffraction is concerned, ADB has been experimentally realized till 8th order by using classical and quantized fields with outstanding results\cite{durr1996pendellosung, vernooy1998high}. This experimental technique has been widely used in Bragg interferometry, with momenta splitting of 102 wavepackets i.e. 102 photons with high splitting efficiency. In such single atom interferometric experiments, fringe visibilities of up to 0.98 ± 0.05 have been recorded, implying high enough splitting efficiencies\cite{chiow2011102}. Therefore, the discussed experimental techniques are well within the workable environment, which compliments our proposal's experimental feasibility. Furthermore, our proposal is expected to have high fidelity because in Bragg regime the interaction times are much longer and the split momenta states are well separated spatially. However, to reduce interaction time errors, ultra-cold $Rb^{85}$ atoms are used most frequently. These atoms are prepared experimentally by magneto-optical trapping, which causes the atom's velocity to spread close to zero \cite{durr1996pendellosung, giltner1995theoretical, durr1998origin}. The diffraction used in the scheme is realized between $Rb^{85}$ and an optical lattice created by a 780nm laser \cite{rempe1990observation, gerry1996proposal}. The experimental conditions for such interactions are, Rb atom should have a mass $M=85amu$ along with the recoil frequency $\omega_r=2.4 \times 10^{4} rad/s$ and coupling constant $\mu=2\pi \times16.4 MHz$. Lastly, the finesse and detunning of the cavity is $4.4 \times 10^{4}$ and $1 GHz$ respectively. Another important feature of this setup is that the cavity in Fock state can easily be realized with a continuous laser beam as a standing wave pattern.
The experimental realization of the atomic quantum channel has been performed successfully in many ways. The recent one is using two individual $Rb^{87}$ atoms, which are trapped individually at different locations approximately 400m apart. To be more precise, it is created through a phenomenon called "Entanglement Swapping", which is conducted over a 700m long optical fiber that connects the two locations \cite{zhang2022device}.

\section{Conclusion} \label{conclusion}
In this paper, we have proposed an experimentally executable Superdense coding scheme, where two parties, Alice and Bob, try to communicate in classical bits via quantum entangled Bell pairs. Our proposal holds a significant value because it is deterministic, state control and its manipulation of atomic states are comparatively easier than its photonic and superconducting \cite{shahmir2023multi} counterparts. Data encoded in the atomic momenta of the atom can travel a larger distance without any decoherence because atomic momenta states are considered resistant to decoherence \cite{kokorowski2001single, ball2008quantum}. It is safe to say that negligible cavity losses, controlled and deterministic systems and decoherent resistant nature, all add to define merits to our experimentally executable proposal.

\bibliographystyle{ieeetr}
\bibliography{sample}

\end{document}